\documentclass[letter]{aa}  
\usepackage{graphicx}
\usepackage{txfonts}
\usepackage{natbib}
\usepackage{subfigure}
\def\sax{SAX~J1818.6$-$1703}

\def\integral{\emph{INTEGRAL}}
\def\xmm{\emph{XMM-Newton}}

\def\swift{\emph{Swift}}
\def\rxte{\emph{RXTE}}
\def\chandra{\emph{Chandra}}
\def\beppo{\emph{BeppoSAX}}
\def\gray{$\gamma$-ray}
\def\cps{counts$\,\mathrm{s}^{-1}$}
\newcommand{\unit}[2]{\mathrm{#1}^{#2}}
\def\ecms{\mathrm{erg}\,\unit{cm}{-2}\,\unit{s}{-1}}
\def\es{\mathrm{erg}\,\unit{s}{-1}}
\def\Msol{M_{\odot}}

\newcommand{\seefig}[1]{(see Fig.~\ref{#1})}

\def\nh{N_{\mathrm{H}}}

\newcommand\ra[3]{#1^{\mathrm{h}}#2^{\mathrm{m}}#3^{\mathrm{s}}}
\newcommand\dec[3]{#1\degr#2\arcmin#3\arcsec}

\begin{document}
   \title{Discovery of an eccentric 30 days period in the supergiant X-ray binary \sax\ with \integral\thanks{Discovery first reported in the 7th \integral\ workshop (Zurita Heras et al., in press). Based on observations with INTEGRAL, an ESA project with instruments and science data centre funded by ESA member states (especially the PI countries: Denmark, France, Germany, Italy, Switzerland, Spain), Czech Republic and Poland, and with the participation of Russia and the USA.}}

   \author{J.A.~Zurita Heras
          \inst{1}
          \and
          S.~Chaty\inst{1}
          }

   \offprints{J.A.~Zurita Heras}

   \institute{Laboratoire AIM, CEA/DSM-CNRS-Universit\'e Paris Diderot,
	IRFU/Service d'Astrophysique, FR-91191 Gif-sur-Yvette, France
	\email{juan-antonio.zurita-heras@cea.fr}, \email{sylvain.chaty@cea.fr}
   }

   \date{Received 2008; accepted XXXX}

 
  \abstract
  {\sax\ is a flaring transient X-ray source serendipitously discovered by \beppo\ in 1998 during an observation of the Galactic centre. The source was identified as a High-Mass X-ray Binary with an OB SuperGiant companion. Displaying short and bright flares and an unusually very-low quiescent level implying intensity dynamical range as large as $10^{3-4}$, the source was classified as a Supergiant Fast X-ray Transient.}
   {The mechanism triggering the different temporal behaviour observed between the classical SGXBs and the recently discovered class of SFXTs is still debated. The discovery of long orbits ($>15$~d) should help to discriminate between emission models and bring constraints.}
   {We analysed archival \integral\ data on \sax. We built short- and long-term light curves and performed timing analysis in order to study the temporal behaviour of \sax\ on different time scales.}
   {\integral\ revealed an unusually long orbital period of $30.0\pm0.2$~d and an elapsed accretion phase of $\sim6$~d in the transient SGXB \sax. This implies an elliptical orbit and constraints the possible supergiant spectral type between B0.5--1I with eccentricities $e\sim0.3-0.4$ (for average fundamental parameters of supergiant stars). During the accretion phase, the source behaved like classical SGXBs. The huge variations of the observed X-ray flux can be explained through accretion of macro-clumps formed within the stellar wind. Our analysis strengthens the model which predicts that SFXTs behave as SGXBs but with different orbital parameters, thus different temporal behaviour.}
  {}

   \keywords{X-ray: binaries -- X-ray: individual: \sax}

   \titlerunning{The 30 days period of \sax}
   \maketitle

%

\section{Introduction}

Supergiant/X-ray binaries (SGXBs) are high-mass X-ray binaries (HMXB) composed of an OB supergiant companion star and a compact object (a neutron star (NS) or a black hole). In SGXBs,  the compact object orbits within a few days (3--15~d) of the massive companion in circular (or slightly eccentric) orbits. Supergiant stars possess a strong stellar wind that is partly captured by the compact object producing an X-ray radiation. This happens either by direct spherical accretion or by Roche-Lobe overflow via an accretion disk on the compact object. As the compact object always orbits within the stellar wind, these systems are persistent X-ray emitting objects and show high variations on short time scales. 

Very few of these systems were known before \integral's launch \citep{Lutovinoval05,Deanal05, Bodagheeal07}. The mission allowed the discovery of several new sources that could be identified as SGXBs \citep[e.g.][]{Walteral06,Chatyal08}. These sources show typical hard X-ray spectra of accreting pulsars and most of them show a strong absorption leading to refer to them as obscured HMXB \citep[e.g.][]{Filliatreal04,Rodriguezal06,Bodagheeal06,Zuritaal06}. Among the new SGXBs, \integral\ has unveiled a new subclass. Indeed, several newly discovered sources were identified as Galactic X-ray sources with a supergiant companion and displaying a transient behaviour \citep[e.g. IGR~J17544$-$2619,][]{Zand05,Pellizzaal06}. The quiescent level in these systems is of the order $10^{32-33}$~ergs$\,\mathrm{s}^{-1}$ \citep[e.g. IGR~J08408$-$4503][]{Gotzal07,Leyderal07}. The X-ray luminosity increases up to $10^{36}$~ergs$\,\mathrm{s}^{-1}$ (as observed in known SGXBs) only during periods of short and luminous flares. The X-ray luminosity remains at a very low level (if not totally absent) during most of the time, except during the flares. These flaring periods last a few hours at most and then the source goes back to an undetectable level of emission \citep[e.g. XTE~J1739$-$302,][]{Smithal06}. Therefore, they received the name of Supergiant Fast X-ray Transient \citep[SFXT,][]{Negueruelaal06}. Besides their transient nature, the spectral features of SFXTs are similar to the previously known persistent SGXBs. 

\beppo\ discovered the new X-ray transient \sax\ on March 11, 1998, during a bright flare \citep{Zandal98}\footnote{All fluxes reported in the literature are summarised in Table~\ref{sumsax}.}. \citet{Zand05} reported its possible association with other fast X-ray transients like XTE~J1739$-$302 and IGR~J17544$-$2619, archetype of SFXTs. Beside detections during deep surveys of the Galactic centre  \citep{Revnivtseval04, Kuulkersal07}, similar short (a few hours) and bright flares of \sax\ were also detected by \integral\  \citep[see Table~\ref{sumsax},][]{Grebeneval05,Sgueraal05,Grebeneval08}. \citet{Grebeneval05} showed that the flares had a complex structure with a short 10--20~min precursor peak and a main 1.5--2~h long peak with different intensities. The hard X-ray spectrum (fitted with simple analytical models such as a power law) was also variable and became harder during the long flares. Using all \integral\ public data available before Mar.~2007, \citet{Walteral07} listed 11 flares with duration of 2--6~ks, and showed that \sax\ was always detected and displayed an intensity dynamic range of 250 between the lowest detection in deep mosaics and the brightest detection in individual pointings. \citet{Zandal06} and \citet{Bozzoal08a} reported a quiescent low level of $\sim(4-8)\times 10^{-12}\ \ecms$ with \chandra\ and \swift, and an unabsorbed 0.5--10~keV flux 3$\sigma$ upper limit of $1.1\times 10^{-13}\ \ecms$ with \xmm\ implying an intensity dynamic range of $10^{3-4}$ as large as in other SFXTs. The \chandra\ spectrum was fitted with an absorbed power law: $\nh=(6.0\pm0.7)\times 10^{22}\ \unit{cm}{-2}$ and $\Gamma=1.9\pm0.3$.

Being similar to other SFXTs and searching a reddened early-type star in optical and infrared catalogues, \citet{Negueruelaal06a} proposed the source 2MASS~J18183790$-$1702479 as the likely counterpart. Confirmed by \chandra, the best X-ray position is R.A.~(2000)~$=\ra{18}{18}{37.9}$ and Dec.~$=\dec{-17}{02}{47.9}$ \citep[$0.6\arcsec$ at 90\% confidence level][]{Zandal06}. The companion spectral type was constrained to a O9--B1 supergiant star \citep{Negueruelaal07a}.

\begin{table}
\caption{Previous detections summary of \sax.}
\label{sumsax}      
\centering          
\begin{tabular}{c c c}
\hline\hline       
Energy range & Flux & phase\\
keV & mCrab & ~\\
\hline                    
2--9$^1$ & 100 & 0.24--0.28 \\
9--25$^1$ & 400 & 0.24--0.28 \\
18--60$^{\star}$ & 4.9 & --\\
18--45$^2$ & 230 & 0.10--0.13\\
45--70$^2$ & 170 & 0.10--0.13\\
20--30$^3$ & 178 & 0.12\\
20--30$^3$ & 185 & 0.16\\
0.5--10$^4$ & $7.5\pm0.1$$^{\ddagger}$ & 0.92--0.96\\
20--60$^{\diamond}$ & $11.1\pm0.9$ & --\\
60--150$^{\diamond}$ & $29.0\pm2.5$ & --\\
18--45$^5$ & 40 & 0.06--0.08\\
0.5--10$^6$ & $<0.11$$^{\ddagger}$ & 0.56--0.59\\
2--10$^7$ & 4.2$^{\ddagger}$ & 0.13--0.17\\
\hline
\end{tabular}
\begin{list}{}{}
\item[$^{\ddagger}$]in unit of $10^{-12}\ \ecms$
\item[$^1$]\beppo, 1998--03--11 \citep{Zandal98}
\item[$^{\star}$]\integral, Aug.--Sept. 2003 \citep{Revnivtseval04}
\item[$^2$]\integral, 2003--09--09 \citep{Grebeneval05}
\item[$^3$]\integral, 2003--10--09 and 10 \citep{Sgueraal05}
\item[$^4$]\chandra, 2006--09--19, exp. 10~ks \citep{Zandal06}
\item[$^\diamond$]\integral, Feb.--Apr.$+$Aug--Oct.  2005 and Feb.--Apr. 2006 \citep{Kuulkersal07}
\item[$^5$]\integral, 2008--04--15 and 16 \citep{Grebeneval08}
\item[$^6$]\xmm, 2006--10--08, exp. 13 ks \citep{Bozzoal08a}
\item[$^7$]\swift, 2008--04--18, exp. 2~ks \citep{Bozzoal08a}
\end{list}
\end{table}

In this study, we used the public \integral\ data in order to look for all the rare and short flares of \sax\ and to study the long-term behaviour of the source, particularly its faint activity. In Sect.~\ref{secObs}, we present the observations, the data analysis, the creation of deep mosaics and the timing analysis. In Sect.~\ref{secRes}, we present the results and discuss the implications in Sect.~\ref{secDis}.

\section{Observations and data analysis}\label{secObs}

Launched in 2002 on a 3-days eccentric orbit, the ESA space mission INTErnational Gamma-Ray Astrophysics Laboratory \citep[\integral,][]{Winkleral03} carries 4 instruments on-board of which 3 are dedicated to the detection of high-energy photons with good angular and spectroscopic resolutions and the last one is an optical $V$-band telescope. In this work, we used data from the hard X-ray and soft \gray\ coded-mask imager IBIS/ISGRI whose main characteristics are a 15~keV--1~MeV energy range, a wide field of view (FOV) of $29\degr\times 29\degr$, and an angular resolution of 12$\arcmin$ \citep{Ubertinial03,Lebrunal03}. 

The observations were divided in individual pointings with usual exposure time of $\sim1-5$~ks. Discarding pointings with exposure times $<600$~s  and source off-axis angles $>14\degr$, we analysed 4003 pointings distributed in 191 revolutions between Feb.~2003 (rev.~{\tt 0046}) and Nov.~2007 (rev. {\tt 0495}). The total exposure time on the source was 9.3~Ms ({\it i.e.}108.4~d) that were unequally distributed along the 4-years observations.

The ISGRI data were reduced using the Offline Scientific Analysis version 7.0 software (OSA7). Individual sky images for each pointing were created in the 22--50 keV energy range. The source count rate was extracted in each image (pointings and mosaics) with the OSA tool {\tt mosaic\_spec} (version 1.7) where the position was fixed to the \chandra\ position and the point-spread function width was fixed to 6$\arcmin$. All detections of the source were considered above the 5$\sigma$ significance level. Once the rare flares localised, we built deep-exposure sky mosaics where the pointings with detections were discarded. The goal was to observe the faint activity of the source and to constrain the lowest hard X-ray flux. We created mosaics for each revolution with exposures between 78--186~ks. Then, we combined consecutive revolutions in which the source was not detected and when the time gap between two revolutions was shorter than 50~d. That resulted in 14 deep mosaics with exposure between 200--1000~ks (except 2 cases with exposures of 60 and 80~ks). 
All the extracted fluxes are gathered to build the light curve of \sax\ \seefig{lcsax}. The 22--50~keV count rate can be converted with the relation $1\ \mathrm{Crab}=177.5\ \mathrm{counts}\,\unit{s}{-1}=9.2\times10^{-9}\ \ecms$. The IJD time unit corresponds to $\mathrm{MJD}=\mathrm{IJD}+51544$.

\section{Results}\label{secRes}

\begin{figure}
\centering
\includegraphics[width=8cm]{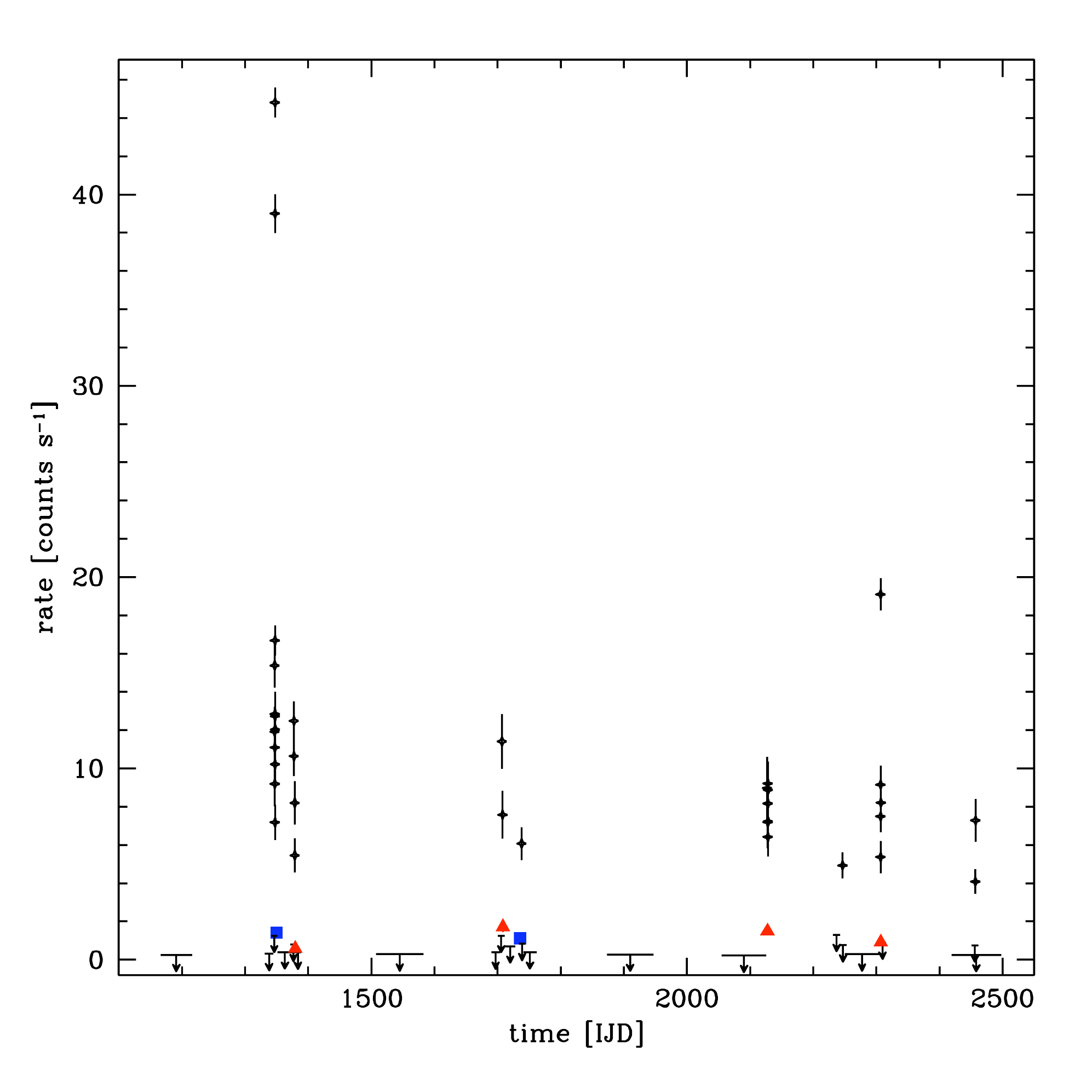}
\caption{22--50~keV light curve of \sax. The data points are described in the text.}
\label{lcsax}
\end{figure}

The source was detected in 32 individual pointings that were distributed in 10 revolutions (crosses in Fig.~\ref{lcsax}). The count rate was mostly between $(4.1-19.1)\pm0.8$~\cps. On Sept.~9, 2003, there were 2 particular very bright flares that reached $44.8\pm0.8$ and $39.1\pm1.0$~\cps\  separated by 2~h \citep[see][]{Grebeneval05}. Discarding all these pointings, we looked for the activity of the source around the flares within the same revolution. The source was only detected in 4 out of 10 revolutions with count rates between $(0.9\pm0.1)-(1.7\pm0.3)$~\cps\ (triangles in Fig.~\ref{lcsax}, and 5$\sigma$ upper limits when the source was not detected). In the other revolution mosaics, the source was only detected in 2 of them with similar count rates (boxes in Fig.~\ref{lcsax}). Both detections occurred before or after a revolution with flares. Beside these detections, we looked for other detections accumulating consecutive revolutions. Even with very long exposure time, the source was not detected with an average 5$\sigma$ upper limit of 0.3~\cps (large 5$\sigma$ upper limits in Fig.~\ref{lcsax}). In total, the source was detected 8 times for a few days at most. Detections 1--2 and 3--4 were both separated by $\sim30$~d, and 6--7 by $\sim60$~d. The source seems to display a periodic recurrent flaring activity separated by $\sim30$~d and lasting for $\sim5$~d. However, flares in several observation periods were missed. 

\begin{figure}
\centering
\includegraphics[width=8cm]{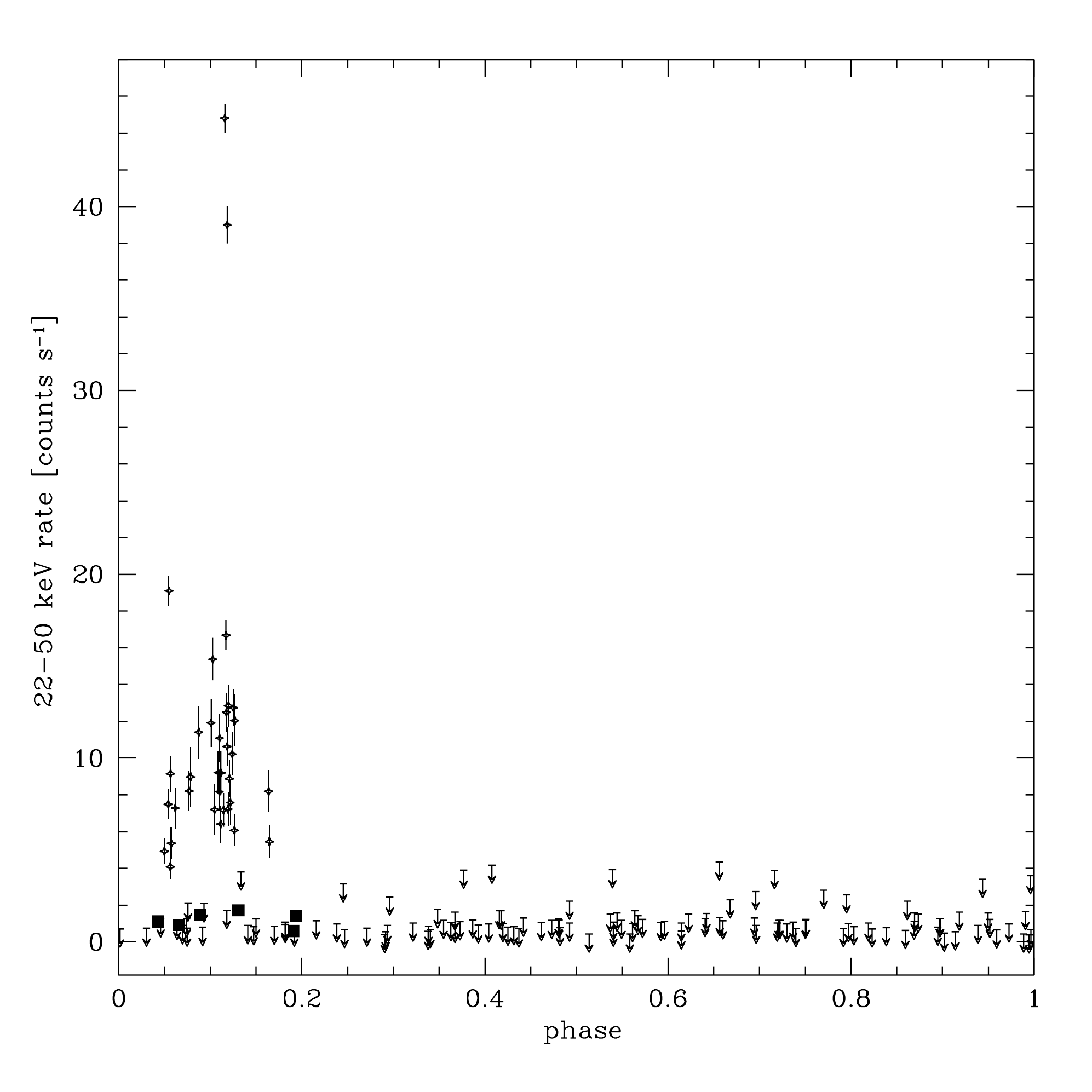}
\caption{22--50~keV folded light curve of \sax\ with a period of $30.0\pm0.2$~d and a time start of IJD~$1133.7365$. Crosses correspond to $\sim2$~ks flares, and boxes and 5$\sigma$ upper-limits correspond to deep mosaics (typical exposure of $\sim100$~ks).}
\label{lcfoldsax}
\end{figure}

To search for the missing flares, we studied the source long-term behaviour using the count rates extracted in pointings and revolution mosaics. In pointings, the sensitivity diminished when the source moved outside of the fully-coded FOV of ISGRI. The 5$\sigma$ level varied between $\sim6-12$~\cps\ when the source off-axis angle varied between $9-14\degr$. As the source was located at off-axis $\geq 9\degr$ in $\sim2/3$ of the pointings, the number of flares below 12~\cps\ was underestimated. In mosaics with exposure  $<20$~ks, the 5$\sigma$ level was usually $>4$~\cps, similar to the count rate of the faint detected flares. Therefore, the Lomb-Scargle periodogram \citep[fast method of][]{Pressal89} was computed using the data of flares and mosaics with exposure $>20$~ks. The best period was $30.0\pm0.2$~d \citep[uncertainty derived using Eq.~14 of][]{Horneal86}. We folded the light curve with this period and a time start of IJD~$1133.7365$ \seefig{lcfoldsax}. All detections fell between phases 0.0--0.2, so an elapsed time of 6~d. Only 31 out of 191 revolutions fell within the flaring periods. \sax\ was detected in 12 of them as described above. We built a mosaic with the other 19 revolutions (exp. 744~ks) and the source was detected with a count rate of $0.47\pm0.06$~\cps, implying a variation factor of 2--4 compared to the average count rates extracted in the 6 revolutions with a detection. Combining revolutions per step of 0.2 phase between 0.2--1.0 (4 mosaics with exposure 1.5--2~Ms each), the source was not detected with a 5$\sigma$ upper limit of $\sim0.2$~\cps.

\section{Discussion}\label{secDis}

As reported above, \sax\ was identified as a SGXB with a spectral type between O9I--B1I. Such systems usually radiate with an average luminosity of $10^{36}\ \es$ with an important variability of factor $\lesssim20$ and in which the compact object orbits around the supergiant companion at a distance $\lesssim2\ R_{*}$. However, the source behaved differently compared to other known SGXBs. For \sax, we found a period of $30.0\pm0.2$~d which can be interpreted as the orbital period of the system. A modulation of activity at a consistent period is found (D.M.~Smith, private communication) in
the \rxte\ Galactic bulge scan data \citep{Markwardtal00}. As observed by \integral, the source only radiated for $\sim6$~d during its orbit. The detections reported by other missions are in good agreement with our ephemeris (see last column in Table~\ref{sumsax}). During the 0.0--0.2 phase range, the average 22--50~keV flux in mosaics varied between $(2-8)\times10^{-11}\ \ecms$. During short flares, the source usually reached fluxes of $\sim(2-6)\times10^{-10}\ \ecms$ and, for very few flares, it even reached $\sim(1-2)\times10^{-9}\ \ecms$. During an \xmm\ observation corresponding to phase $\sim$0.58 ({\it i.e.} near apastron passage), the 0.5--10~keV 3$\sigma$ upper-limit was $<10^{-13}\ \ecms$. Therefore, the compact object must move on an eccentric orbit where it crosses the accretion zone only for 6~d. Only during this phase did \sax\ behave as a SGXB. 

Let us consider the classical accretion model for SGXBs assuming a smooth wind. The semi-major axis $a$ of the system is determined with $a^{3}/P^{2}=G(M_{*}+M_{\mathrm{X}})/4\pi^{2}$ with the period $P$, the stellar mass $M_{*}$, and the mass of the compact object fixed to $M_{\mathrm{X}}=1.4\ \Msol$. The stellar wind velocity of a supergiant can be approximated as $v_{\mathrm{wind}}(r)=v_{\infty}(1-R_{*}/r)^{\beta}$ where $v_{\infty}$ is the terminal velocity, $R_{*}$ the stellar radius, and $\beta=0.8-1.2$. The stellar mass-loss rate $\dot{M}_{\mathrm{wind}}=\dot{M}_{\mathrm{wind}}(L_{*},M_{*},v_{\infty}/v_{\mathrm{esc}},T_{\mathrm{eff}})$ is determined using eq.~12 of \citet{Vinkal00} where $v_{\mathrm{esc}}$ is the stellar escape velocity, and $T_{\mathrm{eff}}$ the stellar effective temperature. Accretion can occur within a radius $R_{\mathrm{accr}}$ determined as $R_{\mathrm{accr}}=2GM_{\mathrm{X}}/(v_{\mathrm{wind}}^2+v_{\mathrm{X}}^2)$ where $v_{\mathrm{X}}$ is the compact object velocity. The accretion rate $\dot{M}_{\mathrm{accr}}$ is determined as $\dot{M}_{\mathrm{accr}}=(R_{\mathrm{accr}}^{2}/4a^{2})\dot{M}_{\mathrm{wind}}$. The luminosity $L_{\mathrm{X}}$ can be approximated as $L_{\mathrm{X}}=\epsilon \dot{M}_{\mathrm{accr}} c^{2}$ with $\epsilon\sim0.2$ the accretion efficiency. Considering typical stellar fundamental parameters of O9I--B1I stars \citep{Martinsal05,Crowtheral06}, the expected accretion rates vary only by factor 2--3 between spectral types. However, they strongly vary with the distance between the star and the compact object with factor $\sim10$ between $r=2-3\ R_{*}$. The expected fluxes when $r=3\ R_{*}$ are consistent with the low flux detected for \sax\ by the different missions. Therefore, we assumed that \sax\ starts to accrete when entering the zone within $r\leq3\ R_{*}$. Still, the size of the accretion zone may vary up to $1\ R_{*}$ as the stellar mass-loss rate can change by a factor of 2--3 in supergiants with identical spectral type \citep{Crowtheral06}.

\begin{figure}
\centering
\includegraphics[width=7.5cm]{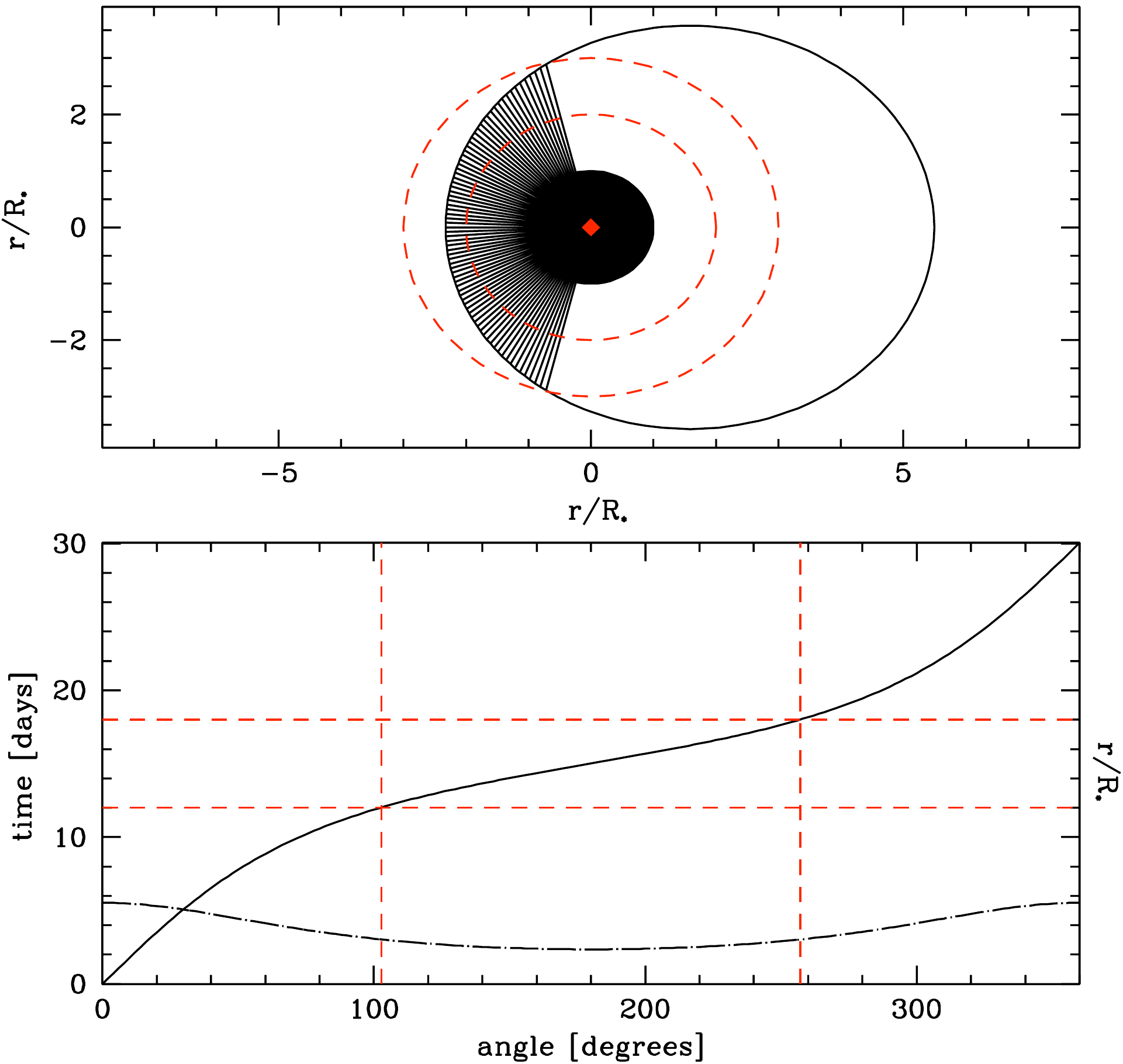}
\caption{{\it Top:} SGXB with elliptical orbit (solid line), B0.5I star (solid circle), and $e=0.41$. Circular dashed lines defined for $r=2$ and $3\ R_{*}$. The accretion zone $(r\leq 3\ R_{*})$ is highlighted. {\it Bottom:} Solid line is the time of the compact object (left axis) along the orbit ($t=0$~d is aphelion); dot-dashed line is the distance star-compact object (right axis) along the orbit; and the light dashed lines determine the accretion zone where $r\leq 3\ R_{*}$ and $t\in t_{\mathrm{accr}}$.}
\label{ellipsesax}
\end{figure}
Under this condition and considering an elliptical orbit\footnote{The Kepler's 2nd law allows to determine $t=t(e,a,\theta (r))$.}, the period and the elapsed time of accretion $t_{\mathrm{accr}}$ derived above implies constrains on the binary system eccentricity $e$. For O9I, O9.5I and B0Ia stars, the maximum elapsed accretion times allowed by the stellar parameters are 2.8, 3.0, and 4.6~d with $e=0.70$, $e=0.69$, and $e=0.58$, respectively, that are inferior to the $t_{\mathrm{accr}}$ derived above. For a B0.5Ia star, the conditions are fulfilled with $e=0.41$ that implies a perihelion and aphelion of 2.3 and 5.5~$R_{*}$ (see Fig.~\ref{ellipsesax}) or $e=0.58$ with $a\in1.6-6.2\ R_{*}$, respectively. For a B1Ia star, the conditions are fulfilled with $e=0.25$ ({\it i.e.} $a\in 2.7-4.5\ R_{*}$) or $e=0.73$ ({\it i.e.} $a\in 1.0-6.2\ R_{*}$). Only B0.5Ia and B1Ia stars seem to fill the accretion condition observed in \sax. Discarding the short and bright flares detections, the X-ray flux varies between $\sim(2-8)\times 10^{-11}\ \ecms$ along the whole $t_{\mathrm{accr}}$. Therefore, the compact object distance $r$ should not vary by a high  factor when crossing the accretion zone, {\it i.e.} low eccentricities are favoured. Therefore, the case of a B0.5--1Ia companion star with a low eccentricity ($e=0.41$ or $0.25$, respectively) are the best solutions to reconcile the wind model and observations. This determination is accurate to 0.5 spectral type since an accretion zone of $4\ R_{*}$ would also give a solution for a B0I star with $e=0.16$.

Short and bright flares were also observed that were 10-100 times higher than the average level during the accretion phase, and $>10^{3-4}$ outside this phase. \citet{Zand05} and \citet{Walteral07} showed that such huge flares in transient SGXBs resulted from the accretion of massive clumps formed within the stellar wind. Indeed, it was shown that instabilities of the radiation line-driven outflows of hot stars generated structure within the stellar wind affecting parameters such as the wind velocity, the outflow density, or the mass-loss rate \citep[][and references therein]{Runacresal02}. Recent studies of supergiant X-ray/UV lines emission have shown that optically-thick macro-clumps must form within the wind of hot supergiant stars \citep[][and references therein]{Oskinovaal07}. Determining average parameters of luminosity, frequency, and duration of flares in SFXTs from \integral\ observations, \citet{Walteral07} derived clump parameters that agreed with this macro-clump scenario developed by \citet{Oskinovaal07}, and predicted that accretion in SGXBs should follow a two-regime behaviour depending on whether the distance between the compact object and the supergiant companion was higher or lower than $\sim 2-3\ R_{*}$. \citet{Negueruelaal08} also expected the same behaviour and showed that the number of clumps that a compact object could statistically accrete within one orbit around a O8-B0I star varies between 1--8 with an orbital radius $a=3-6\ R_{*}$. These values agree with the observation of \sax\ flares, and can explain the observation of flares slightly outside the accretion phase as the flare observed by \beppo.

\begin{acknowledgements}
JAZH thanks Farid Rahoui, J.~Rodriguez, D.M.~Smith and R.~Walter for useful discussions on \sax. The authors also acknowledge the use of NASA's Astrophysics Data System.
\end{acknowledgements}

\bibliographystyle{aa}
\bibliography{../paper/biblio}

\begin{thebibliography}{38}
\expandafter\ifx\csname natexlab\endcsname\relax\def\natexlab#1{#1}\fi

\bibitem[{{Bodaghee} {et~al.}(2007){Bodaghee}, {Courvoisier}, {Rodriguez},
  {Beckmann}, {Produit}, {Hannikainen}, {Kuulkers}, {Willis}, \&
  {Wendt}}]{Bodagheeal07}
{Bodaghee}, A., {Courvoisier}, T.~J.-L., {Rodriguez}, J., {et~al.} 2007, \aap,
  467, 585

\bibitem[{{Bodaghee} {et~al.}(2006){Bodaghee}, {Walter}, {Zurita Heras},
  {Bird}, {Courvoisier}, {Malizia}, {Terrier}, \& {Ubertini}}]{Bodagheeal06}
{Bodaghee}, A., {Walter}, R., {Zurita Heras}, J.~A., {et~al.} 2006, \aap, 447,
  1027

\bibitem[{{Bozzo} {et~al.}(2008){Bozzo}, {Campana}, {Stella}, {Falanga},
  {Israel}, {Rampy}, {Smith}, \& {Negueruela}}]{Bozzoal08a}
{Bozzo}, E., {Campana}, S., {Stella}, L., {et~al.} 2008, The Astronomer's
  Telegram, 1493, 1

\bibitem[{{Chaty} {et~al.}(2008){Chaty}, {Rahoui}, {Foellmi}, {Tomsick},
  {Rodriguez}, \& {Walter}}]{Chatyal08}
{Chaty}, S., {Rahoui}, F., {Foellmi}, C., {et~al.} 2008, \aap, 484, 783

\bibitem[{{Crowther} {et~al.}(2006){Crowther}, {Lennon}, \&
  {Walborn}}]{Crowtheral06}
{Crowther}, P.~A., {Lennon}, D.~J., \& {Walborn}, N.~R. 2006, \aap, 446, 279

\bibitem[{{Dean} {et~al.}(2005){Dean}, {Bazzano}, {Hill}, {Stephen}, {Bassani},
  {Barlow}, {Bird}, {Lebrun}, {Sguera}, {Shaw}, {Ubertini}, {Walter}, \&
  {Willis}}]{Deanal05}
{Dean}, A.~J., {Bazzano}, A., {Hill}, A.~B., {et~al.} 2005, \aap, 443, 485

\bibitem[{{Filliatre} \& {Chaty}(2004)}]{Filliatreal04}
{Filliatre}, P. \& {Chaty}, S. 2004, \apj, 616, 469

\bibitem[{{G{\"o}tz} {et~al.}(2007){G{\"o}tz}, {Falanga}, {Senziani}, {De
  Luca}, {Schanne}, \& {von Kienlin}}]{Gotzal07}
{G{\"o}tz}, D., {Falanga}, M., {Senziani}, F., {et~al.} 2007, \apjl, 655, L101

\bibitem[{{Grebenev} \& {Sunyaev}(2005)}]{Grebeneval05}
{Grebenev}, S.~A. \& {Sunyaev}, R.~A. 2005, Astronomy Letters, 31, 672

\bibitem[{{Grebenev} \& {Sunyaev}(2008)}]{Grebeneval08}
{Grebenev}, S.~A. \& {Sunyaev}, R.~A. 2008, The Astronomer's Telegram, 1482, 1

\bibitem[{{Horne} \& {Baliunas}(1986)}]{Horneal86}
{Horne}, J.~H. \& {Baliunas}, S.~L. 1986, \apj, 302, 757

\bibitem[{{in 't Zand} {et~al.}(1998){in 't Zand}, {Heise}, {Smith}, {Muller},
  {Ubertini}, \& {Bazzano}}]{Zandal98}
{in 't Zand}, J., {Heise}, J., {Smith}, M., {et~al.} 1998, \iaucirc, 6840, 2

\bibitem[{{in't Zand} {et~al.}(2006){in't Zand}, {Jonker}, {Mendez}, \&
  {Markwardt}}]{Zandal06}
{in't Zand}, J., {Jonker}, P., {Mendez}, M., \& {Markwardt}, C. 2006, The
  Astronomer's Telegram, 915, 1

\bibitem[{{in't Zand}(2005)}]{Zand05}
{in't Zand}, J.~J.~M. 2005, \aap, 441, L1

\bibitem[{{Kuulkers} {et~al.}(2007){Kuulkers}, {Shaw}, {Paizis}, {Chenevez},
  {Brandt}, {Courvoisier}, {Domingo}, {Ebisawa}, {Kretschmar}, {Markwardt},
  {Mowlavi}, {Oosterbroek}, {Orr}, {R{\'{\i}}squez}, {Sanchez-Fernandez}, \&
  {Wijnands}}]{Kuulkersal07}
{Kuulkers}, E., {Shaw}, S.~E., {Paizis}, A., {et~al.} 2007, \aap, 466, 595

\bibitem[{{Lebrun} {et~al.}(2003){Lebrun}, {Leray}, {Lavocat}, {Cr{\'e}tolle},
  {Arqu{\`e}s}, {Blondel}, {Bonnin}, {Bou{\`e}re}, {Cara}, {Chaleil}, {Daly},
  {Desages}, {Dzitko}, {Horeau}, {Laurent}, {Limousin}, {Mathy}, {Mauguen},
  {Meignier}, {Molini{\'e}}, {Poindron}, {Rouger}, {Sauvageon}, \&
  {Tourrette}}]{Lebrunal03}
{Lebrun}, F., {Leray}, J.~P., {Lavocat}, P., {et~al.} 2003, \aap, 411, L141

\bibitem[{{Leyder} {et~al.}(2007){Leyder}, {Walter}, {Lazos}, {Masetti}, \&
  {Produit}}]{Leyderal07}
{Leyder}, J.-C., {Walter}, R., {Lazos}, M., {Masetti}, N., \& {Produit}, N.
  2007, \aap, 465, L35

\bibitem[{{Lutovinov} {et~al.}(2005){Lutovinov}, {Revnivtsev}, {Gilfanov},
  {Shtykovskiy}, {Molkov}, \& {Sunyaev}}]{Lutovinoval05}
{Lutovinov}, A., {Revnivtsev}, M., {Gilfanov}, M., {et~al.} 2005, \aap, 444,
  821

\bibitem[{{Markwardt} {et~al.}(2000){Markwardt}, {Swank}, {Marshall}, \& {in't
  Zand}}]{Markwardtal00}
{Markwardt}, C.~B., {Swank}, J.~H., {Marshall}, F.~E., \& {in't Zand}, J.~J.~M.
  2000, in Rossi2000: Astrophysics with the Rossi X-ray Timing Explorer, ed.
  T.~E. {Strohmayer}

\bibitem[{{Martins} {et~al.}(2005){Martins}, {Schaerer}, \&
  {Hillier}}]{Martinsal05}
{Martins}, F., {Schaerer}, D., \& {Hillier}, D.~J. 2005, \aap, 436, 1049

\bibitem[{{Negueruela} \& {Schurch}(2007)}]{Negueruelaal07a}
{Negueruela}, I. \& {Schurch}, M.~P.~E. 2007, \aap, 461, 631

\bibitem[{{Negueruela} \& {Smith}(2006)}]{Negueruelaal06a}
{Negueruela}, I. \& {Smith}, D.~M. 2006, The Astronomer's Telegram, 831, 1

\bibitem[{{Negueruela} {et~al.}(2006){Negueruela}, {Smith}, {Reig}, {Chaty}, \&
  {Torrej{\'o}n}}]{Negueruelaal06}
{Negueruela}, I., {Smith}, D.~M., {Reig}, P., {Chaty}, S., \& {Torrej{\'o}n},
  J.~M. 2006, in ESA SP-604: The X-ray Universe 2005, ed. A.~{Wilson}, 165--170

\bibitem[{{Negueruela} {et~al.}(2008){Negueruela}, {Torrej{\'o}n}, {Reig},
  {Rib{\'o}}, \& {Smith}}]{Negueruelaal08}
{Negueruela}, I., {Torrej{\'o}n}, J.~M., {Reig}, P., {Rib{\'o}}, M., \&
  {Smith}, D.~M. 2008, in American Institute of Physics Conference Series, Vol.
  1010, American Institute of Physics Conference Series, 252--256

\bibitem[{{Oskinova} {et~al.}(2007){Oskinova}, {Hamann}, \&
  {Feldmeier}}]{Oskinovaal07}
{Oskinova}, L.~M., {Hamann}, W.-R., \& {Feldmeier}, A. 2007, \aap, 476, 1331

\bibitem[{{Pellizza} {et~al.}(2006){Pellizza}, {Chaty}, \&
  {Negueruela}}]{Pellizzaal06}
{Pellizza}, L.~J., {Chaty}, S., \& {Negueruela}, I. 2006, \aap, 455, 653

\bibitem[{{Press} \& {Rybicki}(1989)}]{Pressal89}
{Press}, W.~H. \& {Rybicki}, G.~B. 1989, \apj, 338, 277

\bibitem[{{Revnivtsev} {et~al.}(2004){Revnivtsev}, {Sunyaev}, {Varshalovich},
  {Zheleznyakov}, {Cherepashchuk}, {Lutovinov}, {Churazov}, {Grebenev}, \&
  {Gilfanov}}]{Revnivtseval04}
{Revnivtsev}, M.~G., {Sunyaev}, R.~A., {Varshalovich}, D.~A., {et~al.} 2004,
  Astronomy Letters, 30, 382

\bibitem[{{Rodriguez} {et~al.}(2006){Rodriguez}, {Bodaghee}, {Kaaret},
  {Tomsick}, {Kuulkers}, {Malaguti}, {Petrucci}, {Cabanac}, {Chernyakova},
  {Corbel}, {Deluit}, {Di Cocco}, {Ebisawa}, {Goldwurm}, {Henri}, {Lebrun},
  {Paizis}, {Walter}, \& {Foschini}}]{Rodriguezal06}
{Rodriguez}, J., {Bodaghee}, A., {Kaaret}, P., {et~al.} 2006, \mnras, 366, 274

\bibitem[{{Runacres} \& {Owocki}(2002)}]{Runacresal02}
{Runacres}, M.~C. \& {Owocki}, S.~P. 2002, \aap, 381, 1015

\bibitem[{{Sguera} {et~al.}(2005){Sguera}, {Barlow}, {Bird}, {Clark}, {Dean},
  {Hill}, {Moran}, {Shaw}, {Willis}, {Bazzano}, {Ubertini}, \&
  {Malizia}}]{Sgueraal05}
{Sguera}, V., {Barlow}, E.~J., {Bird}, A.~J., {et~al.} 2005, \aap, 444, 221

\bibitem[{{Smith} {et~al.}(2006){Smith}, {Heindl}, {Markwardt}, {Swank},
  {Negueruela}, {Harrison}, \& {Huss}}]{Smithal06}
{Smith}, D.~M., {Heindl}, W.~A., {Markwardt}, C.~B., {et~al.} 2006, \apj, 638,
  974

\bibitem[{{Ubertini} {et~al.}(2003){Ubertini}, {Lebrun}, {Di Cocco}, {Bazzano},
  {Bird}, {Broenstad}, {Goldwurm}, {La Rosa}, {Labanti}, {Laurent}, {Mirabel},
  {Quadrini}, {Ramsey}, {Reglero}, {Sabau}, {Sacco}, {Staubert}, {Vigroux},
  {Weisskopf}, \& {Zdziarski}}]{Ubertinial03}
{Ubertini}, P., {Lebrun}, F., {Di Cocco}, G., {et~al.} 2003, \aap, 411, L131

\bibitem[{{Vink} {et~al.}(2000){Vink}, {de Koter}, \& {Lamers}}]{Vinkal00}
{Vink}, J.~S., {de Koter}, A., \& {Lamers}, H.~J.~G.~L.~M. 2000, \aap, 362, 295

\bibitem[{{Walter} \& {Zurita Heras}(2007)}]{Walteral07}
{Walter}, R. \& {Zurita Heras}, J. 2007, \aap, 476, 335

\bibitem[{{Walter} {et~al.}(2006){Walter}, {Zurita Heras}, {Bassani},
  {Bazzano}, {Bodaghee}, {Dean}, {Dubath}, {Parmar}, {Renaud}, \&
  {Ubertini}}]{Walteral06}
{Walter}, R., {Zurita Heras}, J., {Bassani}, L., {et~al.} 2006, \aap, 453, 133

\bibitem[{{Winkler} {et~al.}(2003){Winkler}, {Courvoisier}, {Di Cocco},
  {Gehrels}, {Gim{\'e}nez}, {Grebenev}, {Hermsen}, {Mas-Hesse}, {Lebrun},
  {Lund}, {Palumbo}, {Paul}, {Roques}, {Schnopper}, {Sch{\"o}nfelder},
  {Sunyaev}, {Teegarden}, {Ubertini}, {Vedrenne}, \& {Dean}}]{Winkleral03}
{Winkler}, C., {Courvoisier}, T.~J.-L., {Di Cocco}, G., {et~al.} 2003, \aap,
  411, L1

\bibitem[{{Zurita Heras} {et~al.}(2006){Zurita Heras}, {de Cesare}, {Walter},
  {Bodaghee}, {B{\'e}langer}, {Courvoisier}, {Shaw}, \& {Stephen}}]{Zuritaal06}
{Zurita Heras}, J.~A., {de Cesare}, G., {Walter}, R., {et~al.} 2006, \aap, 448,
  261

\end{thebibliography}

\end{document}